\documentclass[10pt]{article}

\newcommand{\cudue}{$\mbox{Cu}^{2+}$}

\newcommand{\ntreo}{{N3O1}}
\newcommand{\abinitio}{{\it ab initio}}

\begin{document}

\title{{\it Ab initio} simulations of Cu binding sites\\
 in the N-terminal region of PrP\thanks{Talk given by F.G. at the NIC workshop 'From Computational Biophysics
to Systems Biology 2006', 6-9 June 2006, held at Forschungszentrum J\"ulich, Germany.}}

\author{S. Furlan\thanks{furlan@ge.ismac.cnr.it},\,\, G. La Penna\thanks{lapenna@ge.ismac.cnr.it}\\ \begin{small}CNR,
Institute for Macromolecular Studies, Genova, (Italy)\end{small} \\
\begin{small}and Magnetic Resonance Center, Florence (Italy)\end{small}
\and
F. Guerrieri\thanks{francesco.guerrieri@roma2.infn.it},\,\,
S. Morante\thanks{silvia.morante@roma2.infn.it},\,\,
G.C. Rossi\thanks{giancarlo.rossi@roma2.infn.it}\\
\begin{small}University of Rome ``Tor Vergata'' and INFN Sez. di Roma
``Tor Vergata''\end{small}\\
\begin{small}Via della Ricerca Scientifica - 00133 Roma (Italy)\end{small}}

\maketitle

\begin{abstract}
The prion protein (PrP) binds {\cudue} ions in the octarepeat domain of the N-terminal tail 
up to full occupancy at pH=7.4. Recent experiments show that the HGGG octarepeat subdomain 
is responsible for holding the metal bound in a square planar coordination. By using first 
principle {\abinitio} molecular dynamics simulations of the Car-Parrinello type, the Cu coordination 
mode to the binding sites of the PrP octarepeat region is investigated. Simulations are carried 
out for a number of structured binding sites. Results for the complexes Cu(HGGGW)+(wat), Cu(HGGG) 
and the [Cu(HGGG)]$_2$ dimer are presented. While the presence of a Trp residue and a H$_2$O 
molecule does not seem to affect the nature of the Cu coordination, high stability 
of the bond between Cu and the amide Nitrogens of deprotonated Gly's is confirmed in the case
of the Cu(HGGG) system. For the more interesting [Cu(HGGG)]$_2$ dimer a dynamically entangled 
arrangement of the two monomers, with intertwined N-Cu bonds, emerges. 
This observation is consistent with the highly packed structure seen in experiments 
at full Cu occupancy.
\end{abstract}

\section{Introduction}
The Prion Protein (PrP) is a cell surface glycolipid protein, highly expressed 
in the central nervous system of many mammals. Its physiological r\^ole is still 
unclear, but it has been shown that it can selectively bind {\cudue}. 
A number of Cu binding sites (no less than four) have been identified along the whole protein. 

Mature PrP (comprising a.a. 23-231) has a flexible, disordered, N-terminal (23-120) 
and a globular C-terminal (121-231). The N-terminal domain of human PrP contains four 
repeats of the eight peptides, PHGGGWGQ, from residue 60 to 91, called
\emph{octarepeat}~\footnote{In other species this number may be different. 
For instance, \emph{Bos gaurus}, has 5 octarepeats.}. It has been shown that 
each octarepeat is able to bind one {\cudue} ion~\footnote{Other possible 
binding sites have been found in the C-terminal region.}.

Cu-octarepeat interaction is cooperative, and could possibly have a r\^ole 
in disease related PrP aggregation. Several techniques have been used 
in the study of the octarepeat structural arrangements, including X-ray 
crystallography~\cite{burns} and EXAFS spectroscopy~\cite{morante}. 
Experiments have shown that only the oligopeptide HGGGW is directly involved in {\cudue} coordination.
The crystallographic data shows that the {\cudue} ion is penta-coordinated, 
displaying a square planar equatorial coordination with
three Nitrogens and one Oxygen from the HGGGW peptide, and an axial Oxygen from a water molecule.

The equatorial coordination of the {\ntreo} type is with the $\delta 1$ Nitrogen of the His, 
the deprotonated amide Nitrogens of the two following Gly's and one carbonyl Oxygen from 
the second Gly of the sequence. The axially bound water molecule is kept
in position by a hydrogen bond to the H $\epsilon 1$ of the Trp indole ring.

We have investigated the coordination mode of Cu to the octarepeat via 
{\abinitio} molecular dynamics simulations {\em \`a la} Car--Parrinello. In order to have a system 
of tractable size, we have not considered the whole octarepeat, but only smaller portions
of it, complexed with Cu and water in various combinations. 
More precisely we have investigated the following systems: the Cu(HGGGW)(H$_2$O) complex, 
a box containing Cu(HGGG) and 41 water molecules and a system composed by 
two HGGG oligopeptides, both in the presence and in the absence of an associated pair of {\cudue} ions.

The downsizing of the system here consistent with the experimental findings 
that only the HGGGW portion is directly involved in {\cudue} coordination~\cite{burns, pushie_rauk}.

\section{Methods}
Car--Parrinello MD simulations have been carried out
employing the Quantum-ESPRESSO package~\cite{espresso}. 
We have used Vanderbilt's ultrasoft pseudopotentials and the PBE 
exchange-correlation functional~\cite{pbe}.
Periodic boundary conditions have been imposed on the super-cell, with a minimum 
separation of 5 and 8~{\AA} between replica, for neutral and charged systems, respectively.
The energy cutoff was of 25 Ry, while the hard cutoff for the augmented charge density was of 250 Ry.
All the simulations were spin-restricted. In some particularly interesting cases, namely
the monomer Cu(HGGG) and the dimer [Cu(HGGG)]$_2$, we have performed spin unrestricted simulations, 
with S=1/2 and S=1, respectively.

Every simulation consisted of the following steps.
\begin{enumerate}
\item{
Electronic energy minimization with fixed atomic positions.
}
\item{
Energy minimization with respect to atomic and electronic degrees 
of freedom to attain the equilibrium geometry.
}
\item\label{item:thermalization}{
Two subsequent molecular dynamics simulations of 0.25 ps each at $T=100$ and 200~K, 
respectively~\footnote{A Nos\'e-Hoover thermostat 
at the required temperature was coupled to ionic degrees of freedom.}. }

\item{
Molecular dynamics simulations of about 2 ps at $T=300$~K, 
using the same thermostat as in~\ref{item:thermalization}.
}
\end{enumerate}
Thermalization is necessary to slowly approach room 
temperature and avoid that temperature oscillations obscure the electronic 
properties of the ground state. 
Equations of motion have been integrated using the usual velocity-Verlet algorithm
with a time step of 0.12 fs.
Simulations have been carried out on Linux-clusters using 
8-16 processors, depending on the size of the 
systems~\footnote{For a reference, the monomer Cu(HGGG) comprises 49 atoms and
157 valence electrons.}. 
The CPU time per step was, on average, of 9.6 s when running on 16 nodes at 2.7~GHz.

\section{Results}
Simulations of Cu(HGGGW)(H$_2$O) show that the water molecule in the axial coordination 
site is not bonded to Cu, but rather it is hydrogen-bonded to the Trp side chain.
Indeed we don't see a propensity for a chemical Cu-O(water) bond.
Crystallographic packing effects can explain the presence of water seen in the crystal. 
We have performed simulations in super-cells of increasing size, 
finding that the volatility of the water molecule actually increases. 
We have thus decided to discard the Trp from further simulations. 

Analysis of several trajectories of the complex Cu(HGGG) at $T=300$~K 
show that the {\ntreo} coordination is stable, with the bond between 
Cu and amide Nitrogens from deprotonated Gly's stronger 
than that between Cu and the $\delta 1$ Nitrogen of the His.
The Cu-O bond with the carbonyl group of the second Gly is weak, 
and affected by the greater mobility of the peptide C-terminal.

Simulations of Cu(HGGG) immersed in a box with 41 water molecules confirm 
the picture in which deprotonation of Gly's is favored by the 
greater stability of the bond with Cu, even if at physiological 
pH Gly's are expected to be protonated.

Study of the system [Cu(HGGG)]$_2$ in vacuum shows that, on the time scale of the ps,
formation of a dimer is possible when the tetrapeptides bind copper. Exchange 
of ligands (visible after 0.86 ps at $T=300$~K) between the two Cu ions is seen to keep 
the two tetra-peptides close to each other.
Also for the dimer we observe a preference for an N amidic binding, 
while the bond with the imidazole ring of the His can be broken. 
In spin restricted simulations we observe a rather small Cu-Cu average 
distance (from $2.1$ to $3$ {\AA}) which, in spin unrestricted ones, becomes slightly larger 
($2.4$ to $3.4${\AA}), and thus nearer to experimental observations.
In the absence of Cu, long-range electrostatic and dispersive interactions 
are unable to keep the two (HGGG) close together. We have not yet studied 
the effects of hydrophobic interactions.

\section*{Acknowledgments}
The use of \emph{Cineca} cluster in Bologna, Fermi cluster at \emph{E.~Fermi 
Research Center} in Roma, BEN cluster at \emph{ECT*} in Trento 
and \emph{CERM} cluster in Firenze is gratefully acknowledged. We wish to thank the Organizers 
for the lively atmosphere of the Meeting.

\end{document}